\documentclass[12pt]{article}
\usepackage{psfig, epsfig}
\usepackage{graphicx}

\begin{document}

\begin{center}
\noindent{\Large \bf Unbound Protein-Protein Docking Selections by the DFIRE-based
Statistical Pair Potential} \vskip1cm
\noindent{Song Liu$^1$, Chi Zhang$^1$, and  Yaoqi Zhou$^*$}

\vskip1cm

\noindent{Howard Hughes Medical Institute Center for Single Molecule
Biophysics, Department of Physiology \& Biophysics, State University
of New York at Buffalo, 124 Sherman Hall, Buffalo, NY 14214 }

\noindent{\normalsize $^*$Corresponding Author: Dr. Yaoqi Zhou, Howard
Hughes Medical Institute Center for Single Molecule Biophysics and
Department of Physiology \& Biophysics, State University of New York
at Buffalo, 124 Sherman Hall, Buffalo, NY 14214, Phone: (716)
829-2985, Fax: (716) 829-2344, Email: yqzhou@buffalo.edu } \vskip1cm

\noindent{$^1$ These two authors contribute equally to this work}

\newpage
\begin{abstract}
A newly developed statistical pair potential based on Distance-scaled
Finite Ideal-gas REference (DFIRE) state is applied to
unbound protein-protein docking structure selections. The performance
of the DFIRE energy function is compared to those of the
well-established ZDOCK energy scores and RosettaDock energy function
using the comprehensive decoy sets generated by ZDOCK and
RosettaDock. Despite significant difference in the functional forms
and complexities of the three energy scores, the differences in
overall performance for docking structure selections are small between
DFIRE and ZDOCK2.3 and between DFIRE and RosettaDock. This result is
remarkable considering that a single-term DFIRE energy function was
originally designed for monomer proteins while multiple-term energy
functions of ZDOCK and RosettaDock were specifically optimized for
docking. This provides hope that the accuracy of the existing energy
functions for docking can be improved.
\end{abstract}
\end{center}
\noindent Keywords: potential of mean force, knowledge-based
potential, energy score functions, reference state, binding affinity,
and docking decoys.

\newpage
{\bf\noindent INTRODUCTION}

Docking prediction refers to the prediction of the structure of a
protein-protein complex from the structures of individual subunits.
This is a challenging task because an unbound subunit often changes its
conformation upon binding with its partner (induced fit).  Docking
prediction involves decoy generation and the selection of the
near-native structure from decoys using a filter and/or energy
function. Thus, the success of docking prediction requires an
efficient method that samples near-native conformations and an
accurate energy function that ranks the near-native conformations as
low energy conformations. Advances in sampling methods and energy
functions for docking have been highlighted in several recent
reviews \cite{cama02,Vajda2002,smith02,hal02,janin03,janin03b,mendez03,broo03}.

Various energy functions have been used in docking prediction to
separate near-native structures from other structures. They are
classified into two groups: ``integrated'' and ``edge'' functions
based on whether or not they were used directly in sampling procedures
or applied at the end of sampling procedures \cite{hal02}. Energy
functions are also classified based on the methods used to obtain
them. Physical-based energy functions
\cite{charmm,amber,jorg96,scott99}, derived based on the laws of
physics, have been applied to docking [e.g. DARWIN \cite{DARWIN}, DOT
\cite{Eyck1995}, Hex \cite{Ritchie2000}, Guided Docking
\cite{Bates2003}, TSCF \cite{Komatsu2003}, SmoothDock
\cite{SmoothDock}].  Some docking algorithms use semi-empirical energy
functions that combine various physical terms such as surface
complementarity, van der Waals interaction, generalized Born-surface
area (GB/SA), and hydrogen bonding with optimized weight
factors.  Examples are Dock \cite{Wodak1978,Cherfils1991,wang2003},
ICM-DISCO \cite{Abagyan02}, PPD \cite{Norel1999,Norel2001}, GRAMM
\cite{GRAMM}, FTDOCK \cite{FTDOCK}, 3D-DOCK \cite{3D-DOCK}, AutoDock
\cite{morr98}, Surfdock \cite{Olson1996}, GAPDOCK \cite{Gardiner01},
MolFit \cite{Eisenstein2003,Eisenstein2003b}, BIGGER \cite{BIGGER},
Northwestern DOCK \cite{Shoichet2002}, ZDOCK \cite{rong2003b} and
RosettaDock \cite{gray2003b}.  Still others use statistical energy
functions derived from known protein structures
\cite{zhang97,robert1998,Moont1999,Ponstingl2000,Glaser2001,lu2003,Murphy2003}.
The use of energy functions is often accompanied with clusterization
to incorporate entropic contribution as demonstrated in recent CAPRI
(Critical Assessment of PRedicted
Interaction) \cite{gray2003a,gray2003b,rong2003d}.

Recently, a residue-specific all-atom, distance-dependent potential of
mean-force was extracted from the structures of single-chain proteins
by using a physical state of uniformly distributed points in finite
spheres [distance-scaled, finite, ideal-gas reference (DFIRE) state]
as the zero-interaction reference state \cite{zhou52}. The new energy
function is shown to be one of the best energy functions in selecting
native structures from decoys \cite{zhou52}, predicting
mutation-induced change in stability \cite{zhou52} and loop
conformations \cite{zhou52,zhou62}, and reproducing the partitioning
of hydrophobic and hydrophilic residues within a single
protein \cite{zhou57}.  More importantly, the physical reference state
of ideal gases appears to make the DFIRE energy function physically
more accurate because its performance is largely independent of the
structural database ($\alpha$ or $\beta$ proteins) used for energy
extraction \cite{zhou65}.  Moreover, an initial application of the
DFIRE-based ``monomer'' potential (i.e. the potential extracted from
the structures of single-chain proteins) to protein-protein binding
\cite{zhou61} suggests that the monomer potential is likely to be
useful for docking prediction because it yields a high success rate
for native structure selection in docking decoy sets, discriminates
true dimer from crystal interfaces, and provides an accurate prediction
of protein-protein binding free energies.

In this paper, we further assess the ability of the ``monomer'' DFIRE
energy function to select near-native structures using a large
benchmark of unbound docking decoy sets \cite{rong2003c}. They are the
RosettaDock unbound docking decoy set \cite{gray2003b}, ZDOCK1.3
\cite{rong2002}, ZDOCK2.1 \cite{rong2002}, and ZDOCK2.3
\cite{rong2003b} docking decoy sets. Each docking set contains about
50 protein-protein complexes. We show that the unmodified version of
the DFIRE energy function achieves a success rate in ranking
near-native structures that is comparable to the success rates given
by both ZDOCK and RosettaDock score functions. The implication of this
result is discussed.

{\bf\noindent RESULTS}

{\bf\noindent RosettaDock Unbound Docking Decoy Set}

The DFIRE energy function is tested in the RosettaDock unbound docking
decoy set. As in Ref.~\cite{gray2003b}, the selection capability of a
score function is characterized by the number of structures within the
five lowest energy structures whose root mean squared deviation (rmsd)
values are less than 10\AA~ from the native complex structure ($n_{\rm
rmsd}$) or whose fractions of native residue-residue contacts are
greater than 25\% ($n_{\rm contact}$). Gray {\it et al.} further
defined that a discrimination is successful (a docking funnel is
detected) if $n_{\rm rmsd}$ (or $n_{\rm contact}$) is greater than or
equal to 3.  Table~\ref{tab:rosetta2} compares the performance of the
DFIRE energy function with that of RosettaDock on the docking decoys
of 54 complexes. It shows that the success rate based on $n_{\rm
rmsd}\ge3$ is 32/54 for DFIRE and 34/54 for RosettaDock,
respectively. Similar success rates are obtained if the criterion
$n_{\rm contact}\ge3$ is used. The overall performance of DFIRE
continues to be comparable to RosettaDock when 38 complexes used by
RosettaDock for parameter optimization are removed. Comparable
performance between the two methods is also observed when dividing the
complexes into enzyme/inhibitor, antibody/antigen, and other
complexes. This suggests that the finding is
robust. Figure~\ref{fig:ros} shows several examples in which the DFIRE
energy function produces a ``funnel''-like shape by plotting its
energy score as a function of rmsd from native complex structures.

\goodbreak
{\bf\noindent ZDOCK docking decoy sets}

The DFIRE energy function is applied to docking decoy sets generated
by different versions of ZDOCK. These unbound docking decoy sets
contain 48 protein-protein complexes. In ZDOCK, the success rate is
defined by number of first near-native structures detected within a
given number of energy-ranked structures in the 48 complexes (see
methods). Figure~\ref{fig:success_rate} compares the success rates as
a function of number of energy-ranked structures (or number of
predictions $N_p$) given by DFIRE, ZDOCK1.3, ZDOCK2.0, and ZDOCK2.3.
The results are reported for 16 antibody/antigen complexes, 22
enzyme/inhibitor complexes, 10 other complexes and all 48
complexes. For antibody/antigen complexes, the DFIRE energy function
gives a better success rate than all three versions of ZDOCK except
that at certain intermediate number of predictions (around 10), the
DFIRE energy function gives essentially the same success rate as ZDOCK
1.3 and ZDOCK 2.3. For enzyme/inhibitor complexes, the performance of
the DFIRE energy function continues to be better than that of ZDOCK
2.1 but is only better than that of ZDOCK 1.3 or ZDOCK 2.3 at small
and large $N_p$. For other complexes, the success rates based on top 1
ranking or top 1000 ranking are essentially the same for all four
score functions. At other $N_p$ values, the performance of DFIRE is
essentially the same as that of ZDOCK 2.3, better than that of ZDOCK
1.3, and mixed as compared to ZDOCK 2.1. For all 48 complexes, the
success rate of DFIRE is significantly higher (10\% or more) than that
of ZDOCK 2.1, higher than that of ZDOCK 1.3 for $N_p<5$ or $N_p>30$
and than that of ZDOCK 2.3 for $N_p<4$ or $N_p>30$. The difference
between the results of DFIRE and those of ZDOCK 2.3, however, is
small.

Table~\ref{tab:highest_rank} presents the best rank of near-native
structures given by different methods. In all three decoy sets, DFIRE
increases the ranks of near-native structures for more complexes than
decreases them from the ranks given by different versions of
ZDOCK. More specifically, the ranks given by DFIRE are higher for 23
protein complexes and lower for 12 protein complexes than those given
by ZDOCK 1.3. The corresponding numbers are 27 higher and 9 lower,
relative to ZDOCK 2.1 and 20 higher and 18 lower, relative to ZDOCK
2.3.

Another method to compare different energy score functions is to
compare the number of near-native structures (or number of hits) that are
included within a given number of lowest energy structures (number of
predictions, $N_p$).  Table~\ref{tab:hit_count} compares the number of
near-native structures within the top-1000 decoys given by different
methods in three decoy sets. The application of DFIRE energy function
leads to more protein complexes having a greater number of near-native
structures within the top 1000 decoys. For example, the numbers of
near-native structures given by DFIRE are higher for 22 protein
complexes and lower for 12 protein complexes than those given by ZDOCK
1.3.  The corresponding numbers are 29 higher and 7 lower, relative to
ZDOCK 2.1 and 20 higher and 18 lower, relative to ZDOCK 2.3. The
average number of near-native structures per protein complex given by
DFIRE is higher than ZDOCK 2.1 but is lower than ZDOCK 1.3 and 2.3. We
found that this is mainly caused by relative higher penalty for hard
core overlaps in the DFIRE energy function.  If a softer DFIRE energy
function (see methods) is used, the DFIRE energy function will have a
higher average near-native structures per protein complexes than that
given by three versions of ZDOCK. The softer DFIRE energy function
also further increases the number protein complexes having a greater
number of near-native structures within the top 1000 decoys than those
given by either ZDOCK 1.3, 2.1 or 2.3. We also applied softer DFIRE
energy function to RosettaDock decoy set, but did not find similar
results.  This result indicates that the ZDOCK decoy sets contain
significant van der Waals overlaps whereas the RosettaDock decoy set
has removed those overlaps via minimization.

{\bf\noindent DISCUSSION}

In this paper, we have compared the performance of DFIRE, RosettaDock,
and three versions of ZDOCK in selection of near-native structure from
unbound-proteins docking decoy. The three energy functions were
designed very differently. ZDOCK energy functions were optimally
designed for docking. The shape complementarity was an important
component in ZDOCK. The energy score in ZDOCK 1.3 has three terms:
grid-based shape complementarity, desolvation, and electrostatics. The
energy score in ZDOCK 2.1 uses a pairwise shape complementarity. In
ZDOCK 2.3, the pairwise shape complementarity is further combined with
desolvation and electrostatics.  The RosettaDock energy function, on
the other hand, attempts to include many physical interactions via
physical, empirical, and/or knowledge-based approaches.  The energy
function contains 11 terms that include van der Waals (attractive and
repulsive) interactions, implicit solvation, surface-area solvation,
hydrogen bonding, rotamer probability, residue-residue pair
probability, and electrostatic interactions (short and long-range
attractive and repulsive components). In both ZDOCK and RosettaDock,
weight parameters for different terms were optimized for best
performance. In contrast, the DFIRE energy function only has one
distance-dependent pair potential term that contains no adjustable
parameters (except the energy value for van der Waals core overlaps).
Despite significant difference in three energy functions, the
performance of the DFIRE energy function is comparable to those of
either RosettaDock or ZDOCK 2.3 based on the decoys generated by them.
This is remarkable considering the fact that the DFIRE energy function
was originally designed for monomer proteins.  It remains to be seen
if the performance of DFIRE can be further improved if the DFIRE
energy function is used directly in sampling and minimization (work in
progress).

The result that a single term of statistical pair potential has a
performance similar to multiple-term energy functions provides new
hope for going beyond the existing accuracy of energy functions for
docking. This is because some physical interactions were not taken
into account by the DFIRE energy function. One obvious example is the
multibody hydrogen bonding interaction. Thus, it is possible that
incorporating some terms used in the RosettaDock energy function or
the ZDOCK energy function may further improve the accuracy of the
DFIRE energy function. On the other hand, the matching performance
among three very different energy functions may signal that a
bottleneck in the accuracy of energy function has reached. One
possible source of the error in all three energy functions is implicit
solvation. If this is true, combining additional terms such as
hydrogen bonding and/or surface-accessible solvation with DFIRE will
unlikely make a significant improvement in the accuracy of docking
prediction. Work is in progress to determine which scenario is true.

\begin{figure}[!H]
\vspace{0.5CM}
\centerline{\epsfig{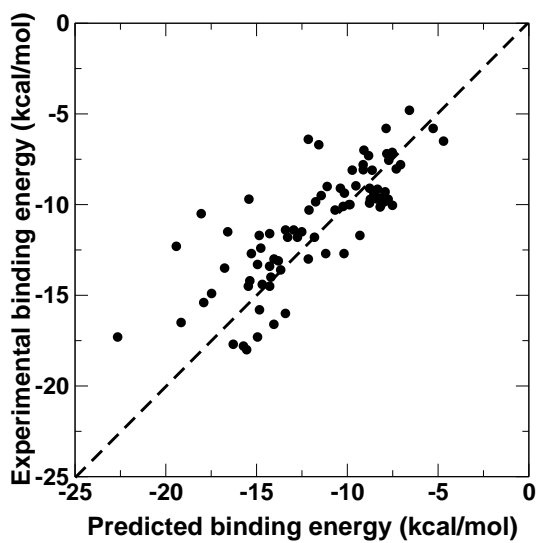}}
\caption{The theoretically predicted binding free energy versus
experimentally measured ones. The line is from linear regression fit
with a correlation coefficient of 0.79, a rmsd of 2.35 kcal/mole. The
dashed line indicates the location if there were a perfect
agreement. }
\label{fig:corr}
\end{figure}

It should be noted that the DFIRE energy function is one of the best
energy functions for predicting the protein-protein (peptide) binding
free energy. Using a combined database of 28 binding free energies
collected by Gray {\it et al.} (2003b) and 69 binding free energies
\cite{zhou61}, the correlation coefficient and the rmsd between
measured binding free energies and that predicted by DFIRE is 0.79 and
2.35 kcal/mole, respectively (See Figure~\ref{fig:corr}).  This
suggests that an accurate prediction of binding free energy does not
guarantee an accurate docking prediction. This further suggests that
the interaction energy missed in the DFIRE energy function only makes
a small contribution to the binding free energy of the native complex
structure but significantly destabilizes other alternative
conformations. This highlights one of the biggest weaknesses of
statistical potentials: they are trained by native structures only.

{\bf\noindent METHODS}

{\bf\noindent DFIRE-based Potential and Soft DFIRE potential}

The derivation of equations, the method for extracting the DFIRE-based
potential using a structure database as well as the resulting
potential have been described or obtained previously \cite{zhou52}.
Here, we give a brief summary for completeness.

The atom-atom potential of mean force $\overline{u}(i,j,r)$ between
atom types $i$ and $j$ that are distance $r$ apart is given
by \cite{zhou52}
\begin{equation}
\overline{u}(i,j,r)=\left\{ \begin{array}{ll} 
-\eta RT \ln\frac{N_{obs}(i,j,r)}{({r\over r_{cut}})^{\alpha}
({\Delta r\over\Delta r_{cut}}) N_{obs}(i,j,r_{cut})}, &r<r_{cut}, \\
0, & r \ge r_{cut},
\end{array} \right.
\label{eq:pot}
\end{equation}
where $\eta=0.0157$, $R$ is the gas constant, $T=300$K, $\alpha=1.61$,
$N_{obs}(i,j,r)$ is the number of $(i,j)$ pairs within the distance
shell $r$ observed in a given structure database, $r_{cut}=14.5$\AA,
and $\Delta r$($\Delta r_{cut}$) is the bin width at $r$($r_{cut}$).
($\Delta r=2$\AA, for $r<2$\AA; $\Delta r=0.5$\AA~ for 2\AA$<r<$8\AA;
$\Delta r=1$\AA~ for 8\AA$<r<$15\AA.) The prefactor $\eta$ was
determined so that the regression slope between the predicted and
experimentally measured changes of stability due to mutation (895 data
points) is equal to 1.0.  The exponent $\alpha$ for the distance
dependence was obtained from the distance dependence for the number of
pairs of ideal gas points in finite spheres (finite ideal-gas
reference state). Residue specific atomic types were used (167 atomic
types) \cite{samu98,lu01}. The number of observed atomic $(i,j)$ pair
within the distance shell $r$ [$N_{obs}(i,j,r)$] was obtained from a
structural database of 1011 non-homologous (less than 30\% homology)
proteins with resolution $<2$\AA~, which was collected by Hobohm {\it
et al.} (1992)
http://chaos.fccc.edu/research/labs/dunbrack/culledpdb.html ).  This
database provides sufficient statistics for most distance bins (except
near the repulsive van der Waals regions). The average number of
observed atomic pairs per bin is 655. The sufficiency of statistics is
also reflected from the fact that the results for structural
discrimination are insensitive to the size of structural database
\cite{zhou52} or the type of structural database \cite{zhou65} used to
generating the potential . The potential $\overline{u}(i,j,r)$ is set
to 10$\eta$ if $N_{obs}(i,j,r)=0$. For a soft-DFIRE energy function,
the value is set to 2$\eta$.

{\bf\noindent Binding Free Energy and Structure Selections from
Docking Decoys}

The total atom-atom potential of mean force, $G$, for each structure
is given by
\begin{equation}
G={1\over2}\sum_{i,j} \overline{u}(i,j,r_{ij}),
\end{equation}
where the summation is over atomic pairs that are not in the same
residue and a factor of 1/2 is used to avoid double counting of
residue-residue and atom-atom interactions. The binding free energy of
a dimer $AB$ is obtained as follows:
\begin{equation}
\Delta G_{\rm bind}=G_{\rm complex}-(G_{A}+G_{B}).
\label{eq:bind}
\end{equation}
Since the structures of monomers are approximated as rigid bodies and
the residues at the interface contribute most to $\Delta G_{\rm
bind}$, Eq.~(\ref{eq:bind}) can be further simplified to
\begin{equation}
\Delta G_{\rm bind}={1\over2}\sum^{\rm interface}_{i,j}
\overline{u}(i,j,r_{ij}),
\label{eq:bind2}
\end{equation}
where the summation is over any two atoms belong to an ``interacting''
residue pair from different chains at the interface. We follow the
definition, due to Lu {\it et al.} (2003), 
 in which an
interacting residue pair is a pair of residues from different chains
that have at least one pair of heavy atoms within 4.5\AA~of each
other. The binding free energy $\Delta G^{\rm decoy}_{\rm bind}$ is
calculated for each docking decoy and the ranking is based on the
value of calculated binding free energy.

{\bf\noindent Unbound Docking Decoy Sets}

The first decoy set (RosettaDock set) consists of 54 decoy sets
[version 1.0 of Chen-Mintseris-Janin-Weng's benchmark \cite{rong2003c}]
downloaded from the website
http://gr aylab.jhu.edu/docking/decoys/. The decoy sets are generated
by random starting position of unbound monomer components superimposed
on the native bound complex structure, followed by 
RosettaDock protocol to create a diffuse space distribution that
covers a reasonable area (~20 \AA~ radius rmsd) with moderate density
around the native position. Each decoy set has 1000 decoys per protein
complex [For more detailed description, see Gray {\it et al.} (2003a).]

The second decoy sets (ZDOCK decoy sets) consist of 48 protein-protein
complexes [version 0.0 of the benchmark \cite{rong2003c}] downloaded
from the website http://zlab.bu. edu/$\sim$rong/dock/software.shtml.  The
decoy sets are generated using fast Fourier transform (FFT) algorithm
based on three different scoring function developed. They are ZDOCK1.3
that combines grid-based shape complementarity, GSC, with desolvation
and electrostatics (GSC+DE+ELEC) \cite{rong2002}, ZDOCK2.1 with
pairwise shape complementarity (PSC) \cite{rong2003a} and ZDOCK2.3,
with combined PSC, desolvation and electrostatics
(PSC+DE+ELEC) \cite{rong2003b}. That is, we have three different
sub-decoy sets and each sub-decoy set has 2000 decoys per protein
complex.

{\bf\noindent Performance Evaluation}

In RosettaDock unbound decoy set, the rmsd between decoy and native
structure is calculated over the C$_{\alpha}$ atoms of the smaller
docking partner (ligand) in the fixed coordinate frame of the larger
partner (receptor). The native residue-residue contact fraction is
calculated as the fraction of the contacts (residue pairs with at
least one inter-residue heavy atom pairs $<$ 4\AA~) identified in the
native structure that are also present in the decoy structures. The
performance of scoring function is evaluated by the number of energy
funnels formed. The unbound perturbation funnels are quantified by
examining the five lowest DFIRE energy decoys. If at least three of
these structures either have less than 10 \AA~ rmsd from the native
structure or a native residue-residue contact fraction above 25\%, a
successful energy funnel exists for this target. [For more detailed
description of the above criterion, see Gray {\it et al.}
(2003a).]
     
In ZDOCK's docking decoys, the rmsd between decoy and native structure is calculated over the C$_{\alpha}$
atoms of interface residues, which are residue pairs between receptor
and ligand with at least one inter-residue heavy atom pairs $<$
10\AA~.  A hit (near-native structure) is defined as decoy with rmsd
$<$ 2.5\AA~.  The performance of a scoring function is evaluated by
using success rate and hit count, as defined by Rong and
Weng (2003). 
 Success rate is defined as the percentage of
test cases in the 48 targets sets for which at least one hit has been
found within a given number of lowest-energy structures (predictions)
for each test case ($N_p$). Hit count is the average number of hits
(near-native structures) per target within a given $N_p$. Success rate
only relies on the first best rank of hit in each protein-protein
complex decoy set. Hit count characterizes the ability to retain near-native
structures for post-processing within a given number of allowed
candidates.

\noindent{\bf \Large Acknowledgments}

We gratefully thank Professors Jeffrey J. Gray and David Baker
for providing us the RosettaDock docking decoy sets, Professor 
Ziping Weng, Dr. Rong Chen, and Ms. Li Li for ZDOCK decoys. We 
are also indebted to Professor Jeffrey J. Gray for many helpful 
discussion during this work. This work was supported by NIH (R01 
GM 966049 and R01 GM 068530), a grant from HHMI to SUNY Buffalo 
and by the Center for Computational Research and the Keck Center 
for Computational Biology at SUNY Buffalo.

\bibliography{docking5}
\bibliographystyle{plain}

\newpage
\begin{table}[]
\setlength{\tabcolsep}{0.15cm}
\footnotesize
\begin{center}
\caption[]{\baselineskip=20pt {\small Comparison of  Performance  in
RosettaDock unbound perturbation of 54 complexes$^a$.}}
\label{tab:rosetta2}
\vspace{2mm}
\begin{tabular}{lcccccccc} \hline\hline
PDB ID    &1ACB$^b$     &1AVW$^b$ &1BRC$^{b,c}$ &1BRS$^{b,c}$  &1CGI$^b$
          &1CHO$^b$  &1CSE$^b$ &1DFJ$^b$\\ \hline
RosettaDock$^d$   &2/1$^e$&5/5&1/2&4/4 &4/4
          &3/3 &2/0&4/4   \\
DFIRE$^f$ &4/4$^g$ &5/5 &2/2 &5/5 &3/3
          &5/5 &1/2 &3/3 \\ \hline
PDB ID    &1FSS$^b$  &1MAH$^{b,c}$ &1TGS$^{b,c}$ &1UGH$^b$ &2KAI$^{b,c}$
          &2PTC$^b$  &2SIC$^b$ &2SNI$^b$\\\hline
RosettaDock$^d$   &5/5 &5/5&5/5 &5/4 &4/4
          &2/2 &5/5&4/4  \\
DFIRE$^f$ &5/0 &5/5&5/5 &5/3 &4/4
          &0/1 &5/5 &5/5 \\ \hline
PDB ID    &\bf{1PPE}$^{b,c}$  &\bf{1STF}$^{b,c}$  &\bf{1TAB}$^{b,c}$ &\bf{1UDI}$^{b,c}$
          &\bf{2TEC}$^{b,c}$  &\bf{4HTC}$^{b,c}$  &1AHW$^h$      &1BVK$^h$ \\\hline
RosettaDock$^d$   &5/5 &5/5 &5/5&5/5&5/5
          &5/5 &5/5 &5/0 \\
DFIRE$^f$ &5/5  &5/5 &3/3 &5/5&4/5
          &5/5  &2/2 &5/1\\\hline
PDB ID    &1DQJ$^h$   &1MLC$^h$  &1WEJ$^h$  &\bf{1BQL}$^h$  &\bf{1EO8}$^h$   &\bf{1FBI}$^h$
          &\bf{1IAI}$^{h,c}$  &\bf{1JHL}$^h$\\ \hline
RosettaDock$^d$   &2/2 &0/0&0/2&5/5&1/4&3/3
          &0/1 &1/0\\
DFIRE$^f$ &1/1  &0/0 &3/1 &1/1 &0/0 &2/3
          &2/2  &1/1\\\hline
PDB ID    &\bf{1MEL}$^{h,c}$  &\bf{1NCA}$^h$  &\bf{1NMB}$^{h,c}$  &\bf{1QFU}$^h$
          &\bf{2JEL}$^h$  &\bf{2VIR}$^{h,c}$  &1AVZ$^i$  &1MDA$^i$\\\hline
RosettaDock$^d$   &5/5 &5/5&5/5&5/5
          &5/4 &4/1 &0/0 & 3/0\\
DFIRE$^f$ &3/4 &3/3 &0/0&4/4
          &5/5  &3/3 &1/0 & 2/1\\\hline
PDB ID    &1WQ1$^i$  &2PCC$^i$      &\bf{1A0O}$^i$     &\bf{1ATN}$^i$
          &\bf{1GLA}$^i$       &\bf{1IGC}$^i$      &\bf{1SPB}$^i$  &\bf{2BTF}$^i$\\\hline
RosettaDock$^d$   &3/4 &3/1&1/4&5/5
          &1/1 &2/2 &5/5 & 4/4\\
DFIRE$^f$ &4/4 &1/3 &3/1&5/5
          &0/0 &0/0 &5/5&5/4\\\hline
PDB ID    &1BTH$^{j,c}$  &1FIN$^j$      &1FQ1$^j$     &1GOT$^j$
          &\bf{1EFU}$^j$       &\bf{3HHR}$^j$ &{\bf\%Total}$^k$ &{\bf\%Subset}$^l$\\\hline
RosettaDock$^d$   &0/1 &0/0&2/2&0/0
          &0/0 &0/0 &34/32 &13/12\\
DFIRE$^f$ &0/0 &0/0 &3/5&0/0
          &0/0  &1/0 &32/30 &12/12\\
\hline\hline
\end{tabular}
\end{center}
\tiny{ $^a$ Bolded targets are decoys from docking between unbound and
bound structures \cite{rong2003c}. Others are between unbound and
unbound structures. $^b$ The enzyme/inhibitor complexes. $^c$ The
complexes that were not used for optimizing the weighting scores in
the RosettaDock energy function. $^d$ The High-resolution
RosettaDock scoring function \cite{gray2003a,gray2003b}.  $^e$ The
first (second) number in the cell is the number of top 5 decoys with
rmsd$<$10\AA~ (more than 25\% of native residue-residue contact) given
by the RosettaDock scoring function.  $^f$ The DFIRE-based potential
derived from a database of single-chain proteins \cite{zhou52}. 
$^g$ The first (second) number in the cell is the number of top 5
decoys with rmsd$<$10\AA~ (more than 25\% of native residue-residue
contact) given by DFIRE scoring function. $^h$ The antibody/antigen
complex.  $^i$ Other complexes.  $^j$ Difficult targets. $^k$ The
success rate based on the number of targets that have greater than or
equal to three rmsd$<$10\AA~ (or more than 25\% native contact decoys)
ranked in top 5 as in Ref. \cite{gray2003b}.  $^l$ The success rates of
the independent subset for the complexes that were not used in weight
optimization. }
\end{table}

\newpage
\begin{table}[!h]
\scriptsize
\begin{center}
\caption[]{Highest rank of hits in ZDOCK docking decoy$^a$}
\vspace{1mm}
\begin{tabular}{lllllll} \hline\hline
 & \multicolumn{2}{c}{\underline{ZDOCK1.3 DECOYS$^b$}}  &\multicolumn{2}{c}{\underline{ZDOCK2.1 DECOYS$^c$}} &\multicolumn{2}{c}{\underline{ZDOCK2.3 DECOYS$^d$}} \\
PDB ID &GDE$^e$ &DFIRE$^f$ &PSC$^g$ &DFIRE$^f$ &PDE$^h$ &DFIRE$^f$ \\ \hline
1MLC
&134 &64  
&1396&141   
&128 &146     \\ 
1WEJ
&1940 &159
&1106 &406
&183  &36    \\ 
1AHW
&11 &10
&26 &5
&7  &4    \\ 
1DQJ
&--   &--
&1341 &312 
&--   &--  \\ 
1BVK
&--   &--
&974  &1386 
&821  &1239    \\ 
1FBI$^i$
&561 &812 
&1786&1619
&642 &1418   \\ 
2JEL$^i$
&--  &-- 
&112 &214 
&233 &1030   \\ 
1BQL$^i$
&4   &24
&172 &28
&13  &31  \\ 
1JHL$^i$
&--  &--
&404 &116
&333 &51   \\ 
1NCA$^i$
&211 &90 
&2   &14
&1   &152 \\ 
1NMB$^i$
&1108 &329   
&693  &215   
&135  &28    \\ 
1MEL$^i$
&9  &1
&12 &1
&3  &1  \\ 
2VIR$^i$
&--  &--
&476 &125 
&1101&315 \\ 
1EO8$^i$
&--  &--
&--  &--
&1497&141  \\ 
1QFU$^i$
&606  &48
&407  &1
&388  &1\\ 
1IAI$^i$
&905  &102   
&-- &--  
&997  &299   \\
\\

1CGI
&3 &4 
&4 &7 
&4 &78 \\ 
1CHO
&22 &1 
&1  &1 
&3  &1  \\ 
2PTC
&65    &14 
&1655  &715  
&193   &15  \\ 
1TGS
&5  &1 
&3  &4 
&3  &9    \\ 
2SNI
&169 &331 
&-- &-- 
&1262  &913  \\ 
2SIC
&2   &126 
&241 &13 
&11  &95  \\ 
1CSE
&3  &30 
&1537  &37  
&198 &9  \\ 
2KAI
&1772 &1044 
&1399  &264
&388  &212   \\ 
1BRC
&52  &25
&173 &13
&24  &13   \\ 
1ACB
&3   &25 
&25  &33  
&18  &70   \\ 
1BRS
&1019 &244
&61   &44
&65   &131   \\ 
1MAH
&9    &12 
&849  &39
&24   &41   \\ 
1UGH
&14   &30  
&305  &316   
&8    &22      \\ 
1DFJ
&2  &48   
&37 &14  
&1  &26       \\ 
1FSS
&1066  &113   
&731   &259   
&50    &97      \\ 
1AVW
&704  &37   
&45   &16  
&3    &16      \\ 
1PPE$^i$
&1   &1 
&1   &1 
&1   &1     \\ 
1TAB$^i$
&--  &--  
&65  &6  
&79  &5     \\ 
1UDI$^i$
&198 &33 
&31  &1 
&5   &1      \\ 
1STF$^i$
&1   &1 
&1   &1  
&1   &1    \\ 
2TEC$^i$
&1   &2 
&1   &1  
&1   &1     \\ 
4HTC$^i$
&2  &2  
&1  &1  
&3  &3     \\
\\
2PCC
&702  &234
&-- &-- 
&-- &-- \\
1WQ1
&131  &82  
&5    &3 
&15   &5 \\ 
1AVZ
&-- &--    
&-- &--    
&-- &--    \\ 
1MDA
&-- &--    
&-- &--    
&-- &--    \\ 
1IGC$^i$
&--  &--  
&22  &239   
&153 &452   \\ 
1ATN$^i$
&13   &2 
&360  &127  
&7    &7 \\ 
1GLA$^i$
&214  &53 
&--  &--
&--  &--\\ 
1SPB$^i$
&1  &2 
&1  &5 
&1  &13 \\ 
2BTF$^i$
&27 &1 
&32 &1 
&2  &1 \\ 
1A0O$^i$
&619 &108 
&833 &139 
&284 &218 \\
\\ 

Ratio$^j$
& --  &12/23
& --  &9/27 
& --  &18/20\\ 

Top 1$^k$
&4&6
&6&9
&6&7\\ 

 \hline \hline
\end{tabular}
\label{tab:highest_rank}
\end{center}
\tiny{
$^a$ The 1JTG decoy set is not available (See
http://zlab.bu.edu/$\sim$rong/dock/software.shtml). Hits are defined as
docked structures with interface rmsd$\le$2.5\AA~ from the crystal
complexes. There are 2000 decoys for each target. $^b$Decoys generated
by ZDOCK1.3 \cite{rong2002}.  $^c$Decoys generated by ZDOCK2.1
\cite{rong2003a}.  $^d$Decoys generated by ZDOCK2.3
\cite{rong2003b}.  $^e$ZDOCK1.3(GDE) \cite{rong2002}.  $^f$The
DFIRE-based potential \cite{zhou52}.  $^g$ZDOCK2.1(PSC)
\cite{rong2003a}.  $^h$ZDOCK2.3(PDE) \cite{rong2003b}.  $^i$ Decoys
from docking between unbound and bound structures.  $^j$ The first
(second) number is the number of targets whose ranks given by DFIRE
are lower (higher) than that given by the ZDOCK scoring
function. $^k$The number of targets whose near-native structures are
scored as top 1. 
}
\end{table}

\newpage
\begin{table}[!h]
\scriptsize
\begin{center}
\caption[]{Hit scored in top-1000 of ZDOCK docking decoy$^a$}
\vspace{1mm}
\begin{tabular}{llllllllll} \hline\hline
 & \multicolumn{3}{c}{\underline{ZDOCK1.3(GDE) DECOYS$^b$}}  &\multicolumn{3}{c}{\underline{ZDOCK2.1(PSC) DECOYS$^c$}} &\multicolumn{3}{c}{\underline{ZDOCK2.3(PDE) DECOYS$^d$}} \\ 
PDB ID &GDE$^e$ &DFIRE$^f$ &SoftDFIRE$^g$
 &PSC$^h$ &DFIRE$^f$ &SoftDFIRE$^g$ &PDE$^i$ &DFIRE$^f$ &SoftDFIRE$^g$\\ \hline
1MLC
&11 &11  &14
&0  &3   &3
&11 &8   &9  \\ 
1WEJ
&0  &1   &0
&0  &3   &4
&11 &19  &21    \\ 
1AHW
&44 &63  &57
&18 &21  &26
&50 &56  &59    \\ 
1DQJ
&-- &--  &--
&0  &1   &0
&-- &--  &--  \\ 
1BVK
&-- &--  &--
&1  &0   &0
&1  &0   &1   \\ 
1FBI$^j$
&1 &2   &2
&0 &0   &1
&1 &0   &3  \\ 
2JEL$^j$
&--  &--    &--
&33  &21    &43
&17  &0     &19 \\ 
1BQL$^j$
&73  &68    &85
&6   &12    &11
&46  &30    &51  \\ 
1JHL$^j$
&--  &--    &--
&7   &10    &6
&5   &9     &7  \\ 
1NCA$^j$
&6   &9     &7
&43  &34    &41
&55  &25    &47 \\ 
1NMB$^j$
&0 &3   &2
&3 &5   &0
&7 &7   &2 \\ 
1MEL$^j$
&19 &30 &32
&36 &47 &52
&52 &58 &71  \\ 
2VIR$^j$
&-- &--  &--
&1  &3   &0
&0  &3   &0\\ 
1EO8$^j$
&--  &--   &--
&--  &--   &--
&0   &2    &1 \\ 
1QFU$^j$
&4  &4     &4
&2  &10    &10
&10 &18    &18\\ 
1IAI$^j$
&1  &3   &3
&-- &--  &-- 
&1  &3   &2\\ 
\\ 

1CGI 
&43 &52 &70  
&29 &37 &42   
&50 &42 &53 \\ 
1CHO   
&53 &85 &76 
&54 &65 &65      
&68 &85 &79  \\ 
2PTC 
&38 &43 &61
&0  &2  &1    
&28 &25 &33  \\ 
1TGS 
&60 &72 &86  
&87 &85 &106  
&79 &82 &105   \\ 
2SNI 
&34 &19 &46 
&-- &-- &--     
&0  &1  &0 \\ 
2SIC 
&96 &53 &105 
&10 &20 &16    
&38 &28 &30  \\ 
1CSE 
&61 &30 &48   
&0  &3  &0    
&15 &22 &8   \\ 
2KAI 
&0  &0 &1 
&0  &2 &2   
&3  &13&14   \\ 
1BRC 
&9  &17 &20  
&13 &16 &16     
&35 &47 &49   \\ 
1ACB 
&154 &120 &136  
&21  &25  &32     
&54  &51  &54  \\ 
1BRS 
&0  &1   &3 
&20 &25  &32     
&14 &14  &16  \\ 
1MAH 
&45 &40  &51   
&3  &6   &6     
&22 &24  &26   \\ 
1UGH 
&36 &43  &53  
&2  &3   &4    
&14 &14  &20    \\ 
1DFJ 
&36 &9   &43    
&13 &10  &15       
&44 &14  &45     \\ 
1FSS 
&0  &2   &2
&1  &4   &4    
&11 &10  &13    \\ 
1AVW  
&2  &2   &2 
&18 &21  &27    
&39 &34  &49    \\ 
1PPE$^j$ 
&257 &143 &230   
&215 &198 &218      
&325 &233 &296    \\ 
1TAB$^j$ 
&--  &--  &-- 
&31  &40  &29   
&28  &39  &31   \\ 
1UDI$^j$ 
&28 &29 &32
&13 &13 &16   
&26 &30 &35     \\ 
1STF$^j$   
&140 &120 &143 
&37  &39  &42  
&67  &67  &77   \\ 
2TEC$^j$ 
&191 &138 &168 
&64  &69  &57  
&151 &127 &109    \\ 
4HTC$^j$ 
&53 &54  &65 
&40 &47  &52  
&36 &44  &52   \\ 
\\\vspace{-0.2CM}
2PCC
&2 &5   &5
&-- &-- &--
&-- &-- &--\\ 
1WQ1
&5  &8  &10
&21 &17 &22
&40 &29 &46\\ 
1AVZ
&-- &--    &--
&-- &--    &--
&-- &--    &--\\ 
1MDA
&-- &--    &--
&-- &--    &--
&-- &--    &--\\ 
1IGC$^j$
&-- &--  &--
&4  &4   &1
&2  &2   &1\\ 
1ATN$^j$
&30 &44 &40
&1  &1  &0
&16 &23 &2\\ 
1GLA$^j$
&8  &19 &19
&--  &--   &--
&--  &--   &--\\ 
1SPB$^j$
&85  &84 &102
&59  &59 &73
&98  &71 &105\\ 
2BTF$^j$
&15 &14 &14
&11 &13 &13
&28 &33 &30\\ 
1A0O$^j$
&4 &9 &6
&1 &2 &2
&4 &3 &2\\
\\ 

Ratio$^k$
& --  &12/22&5/30
& --  &7/29 &9/30
& --  &18/20&10/29\\ 

Average$^l$
&34.25&30.19&38.40
&19.13&20.73&22.71
&33.38&30.10&35.23\\ 

\\ \hline \hline
\end{tabular}
\label{tab:hit_count}
\end{center}
\tiny{
$^a$ The 1JTG decoy set is not available (See
http://zlab.bu.edu/$\sim$rong/dock/software.shtml). Hits are defined
as docked structures with interface rmsd$\le$2.5\AA~ from the crystal
complex. There are 2000 decoys for each target. $^b$Decoys generated
by ZDOCK1.3 \cite{rong2002}.  $^c$Decoys generated by ZDOCK2.1
\cite{rong2003a}.  $^d$Decoys generated by ZDOCK2.3
\cite{rong2003b}.  $^e$ZDOCK1.3(GDE) \cite{rong2002}.  $^f$The
DFIRE-based potential \cite{zhou52}.  $^g$The Soft-DFIRE potential.
$^h$ZDOCK2.1(PSC) \cite{rong2003a}.  $^i$ZDOCK2.3(PDE)
\cite{rong2003b}.  $^j$Decoys from docking between unbound and bound
structures.  $^k$The first (second) number are the number of targets
whose number of hits given by DFIRE/SOFT-DFIRE are lower (higher) than
that given by the ZDOCK scoring function. $^l$ The average number of
hits over 48 targets.
}
\end{table}

\newpage

\newcommand{\asgltwo}[2]{\begin{tabular}{c@{\quad}@{\quad}@{\quad}@{\quad}c}
  \mbox{\epsfig{file=#1,width=7cm}}
&  \mbox{\epsfig{file=#2,width=7cm}}
\end{tabular}}

\begin{figure}[!H]
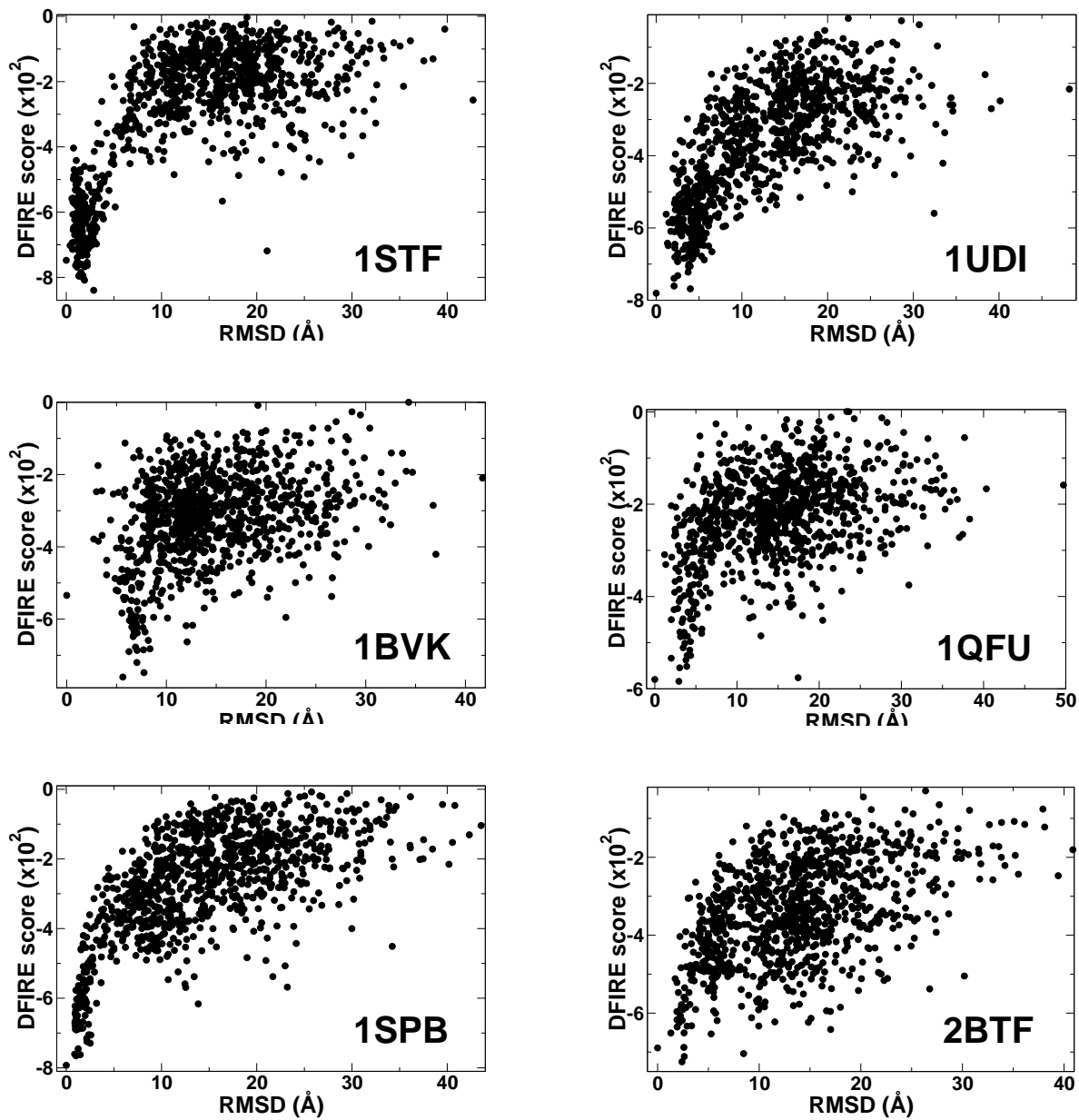

\centerline{\asgltwo
{1STF.eps}
{1UDI.eps} }

\vspace{0.5CM}
\centerline{\asgltwo
{1BVK.eps}
{1QFU.eps}
}
\vspace{0.5CM}
\centerline{\asgltwo
{1SPB.eps}
{2BTF.eps}
}
\caption{Scatter plots of the DFIRE score {\it versus}
rmsd of RosettaDock decoy from the native structure (based on C$_{\alpha}$ atoms).
Results of two proteins (1stf at the top left and 1nca at the top right)
from the enzyme/inhibitor complexes, two proteins (1bvk at the middle
left and 1qfu at the middle right) from the antibody/antigen complexes,
and two proteins (1spb at the bottom left and 2btf at the bottom right)
from the other complexes are shown. }
\label{fig:ros}
\end{figure}

\newpage
\newcommand{\asglfour}[3]{\begin{tabular}{c@{\quad}@{\quad}c@{\quad}@{\quad}c}
  \mbox{\epsfig{file=#1,width=5cm}}
&  \mbox{\epsfig{file=#2,width=5cm}}
&  \mbox{\epsfig{file=#3,width=5cm}}
\end{tabular}}

\begin{figure}[!H]
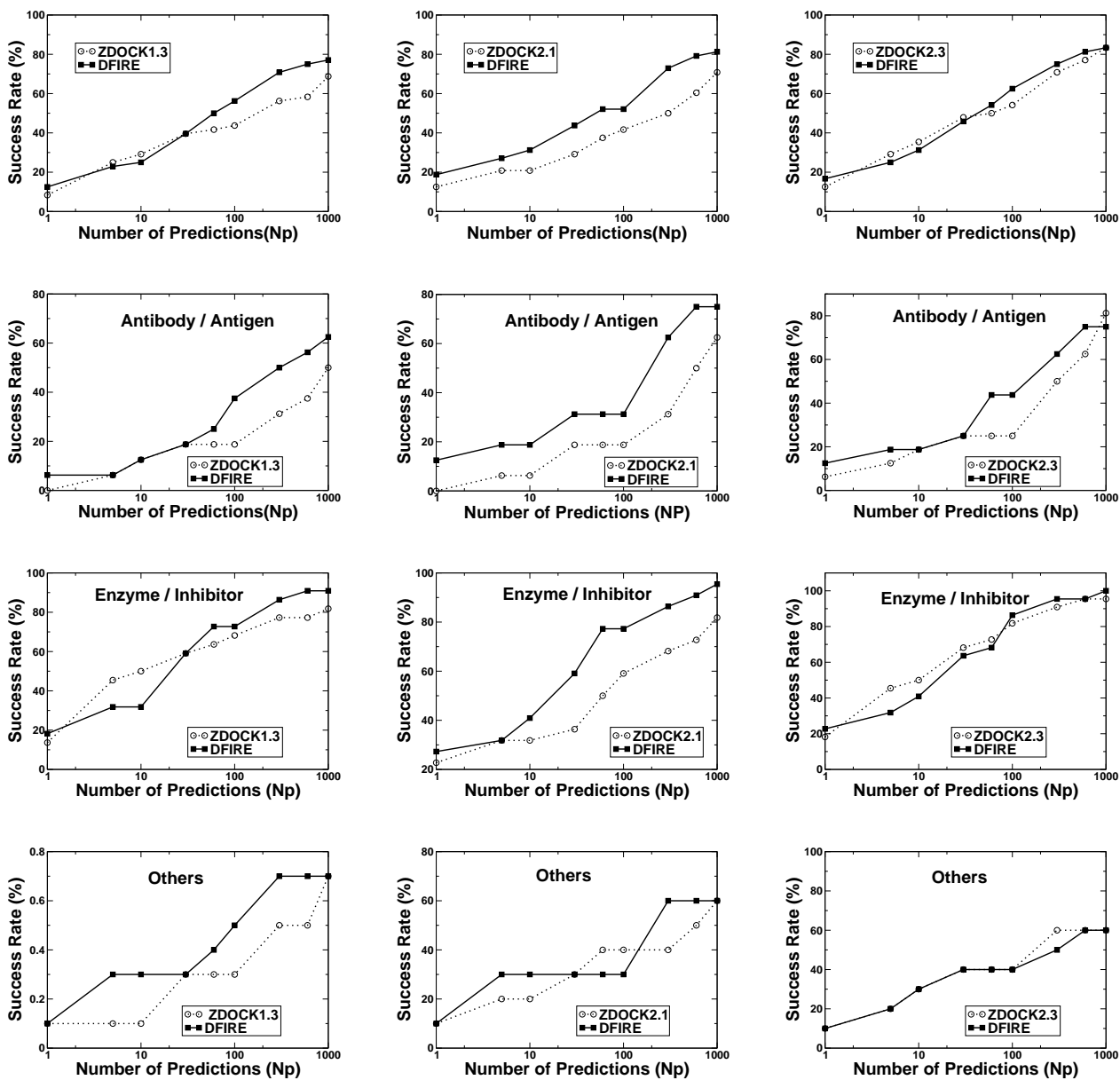

\vspace{0.5CM}
\centerline{\asglfour
{sr1.3.eps}
{sr2.1.eps}
{sr2.3.eps}
}

\vspace{0.5CM}
\centerline{\asglfour
{sr1.3.a.eps}
{sr2.1.a.eps}
{sr2.3.a.eps}
}

\vspace{0.5CM}
\centerline{\asglfour
{sr1.3.e.eps}
{sr2.1.e.eps}
{sr2.3.e.eps}
}

\vspace{0.5CM}
\centerline{\asglfour
{sr1.3.o.eps}
{sr2.1.o.eps}
{sr2.3.o.eps}
}

\caption{The performance of ZDOCK1.3 (left), ZDOCK2.1 (middle),
ZDOCK2.3 (right) are compared to that of DFIRE according to success
rates as a function of number of predictions (number energy-ranked
structures) in 16 antibody-antigen decoy sets (top), 22
enzyme-inhibitor decoy sets (middle up) and 10 other complexes decoy
sets (middle bottom) and 48 overall decoy sets (bottom)}
\label{fig:success_rate}
\end{figure}

\end{document}